\documentclass[12pt]{article}
\usepackage{graphicx}

\begin{document}

\begin{center}
{\bf Higgs Boson as a Dilaton}\\
\vspace{0.5cm}
M.G. Ryskin, A.G. Shuvaev\\

Petersburg Nuclear Physics Institute, Gatchina, S.Petersburg,
188300,
Russia\\

\end{center}

\begin{abstract}
We study possible phenomenological consequences of the recently proposed
new approach to the Weinberg-Salam model. The electroweak theory is
considered as a gravity and the Higgs particle is interpreted in it as a dilaton,
without the usual potential of interaction in the Higgs sector.
We have taken as a test the process of photons pair production,
$e^+ + e^- \to Z + \gamma + \gamma$. In the framework of new formulation
this reaction is mediated in the lowest order by the dilaton. The cross section
is found to be rather small.
\end{abstract}

\section{Introduction}
Recently, a new formulation of the Weinberg-Salam model has been suggested
\cite{Chernodub:2008rz,Faddeev:2008qc,Faddeev:2006sw,Chernodub:2007bz}
(the similar ideas are discussed also in papers
\cite{Diakonov:2001xg,Foot:2008tz,Majumdar:2009yw}).
A novel feature of this approach is the interpretation of the electroweak theory
in terms of a gravity theory with the Higgs field as a dilaton.

The bosonic sector of the standard electroweak theory given by
the Lagrangian including Yang-Mills triplet $W_\mu^a$, abelian vector field $Y_\mu$
and complex scalar Higgs dublet $\Phi=(\phi_1,\phi_2)$\footnote{Our notations are a bit
different from those used in the papers \cite{Chernodub:2008rz} - \cite{Chernodub:2007bz}},
\begin{equation}
\label{EW}
{\cal L}_{WS}\,=\,-\frac 14\,G_{\mu \nu}^2(W)\,-\,\frac 14 F_{\mu \nu}^{\,2}(Y)\,
+\,|D_\mu \Phi|^2\,-\,\,\mu^2|\Phi|^2 \,-\,\lambda |\Phi|^4,
\end{equation}
can be equivalently reformulated as an effective gravity\footnote{The Higgs mass $\mu=0$
in ${\cal L}_{WS}$, the case $\mu \neq 0$ is discussed in the next section.},
\begin{equation}
\label{LWS}
{\cal L}_{WS}\,=\,\sqrt{-{\cal G}}\,\bigl[-(M_p^2R + \lambda)\,+
\,{\cal L}_{M}\bigr].
\end{equation}
Here $R$ is the scalar curvature, $M_p$ plays the role of a Plank mass and the second term
is the matter Lagrangian,
\begin{eqnarray}
{\cal L}_{M}\,&=&\,-\frac 14\,{\cal G}^{\mu\rho}{\cal G}^{\nu\sigma}
G_{\mu\nu}G_{\rho\sigma}\,
-\,\frac 14\,{\cal G}^{\mu\rho}{\cal G}^{\nu\sigma}F_{\mu\nu}F_{\rho\sigma}\,
+\,\frac 12 M_Z^2\,{\cal G}^{\mu\nu}Z_\mu Z_\nu\, \nonumber \\
\label{matter lagr}
&+&\,M_W^2\,{\cal G}^{\mu\nu}W_\mu^+ W_\nu^-
\end{eqnarray}
including massive vector bosons $Z$ and $W^\pm$ and massless photon $A_\mu$. Lagrangian
${\cal L}_{M}$ does not contain Higgs boson, which is interpreted in this approach
as a dilaton and gives rise to the first, gravity, term. The metric tensor is always taken
to be conformally flat,
\begin{equation}
\label{rh}
{\cal G}_{\mu\nu}\,=\,\frac{\rho^{\,2}}{m^2}\,\eta_{\mu\nu},
\end{equation}
with flat Minkowski metric $\eta_{\mu\nu}=(+---)$, where $\rho$ is the modulus of
the Higgs field, $\rho^2=|\Phi|^2$, and $m$ is the arbitrary parameter having dimension
of mass. It provides a scale through which all other mass parameters are expressed,
\begin{equation}
\label{mscale}
\frac 14\, g^2 m^2\,=\,M_W^{\,2},~~~ \frac 12 \,(g^2+g^{\prime \,2})m^2\,=\,M_Z^2,~~~
M_p^{\,2}\,=\,\frac{1}{6}\,m^2.
\end{equation}
The kinetic part of the dilaton field, $(\partial \rho)^2$,
turns into the scalar curvature in the gravity action\footnote{There is a possible
ambiguity related to the analytical continuation from
Euclidean space, where the theory is originally formulated, to Minkowski space.
It could change the sign in front of the curvature term in the gravity action
\cite{Chernodub:2008rz}. However this sign seems not to be especially important
for our purposes.},
$$
\frac{m^2}6\sqrt{-{\cal G}}\,R\,=\,-\bigl(\partial \rho \bigr)^2\,+\,
\mbox{total derivatives}.
$$
The vector bosons acquire finite masses for $m \neq 0$ in the absence of the usual
Higgs mechanism based on a symmetry breaking. There is no strict relation between
the value of $m$ and scalar potential parameter $\lambda$ or Higgs mass $\mu$,
the non-zero masses can be generated in this interpretation even for $\lambda=0$.
Moreover, the dilaton, which replaces the Higgs particle in the effective gravity
action (\ref{LWS}), has always zero mass. These properties mark an essential difference
to conventional approaches that underly the Standard Model. This is the reason to study
possible phenomenological consequences following from this theory.

\section{Comments on perturbation theory}
Before going to phenomenological aspects we should comment a little on the perturbation
theory since we are about to exploit it for the effective gravity. To develop the perturbation
theory we need first to derive a functional integral. The functional measure is originally
given by the product of appropriate measures for all fields in the standard model (bosonic part),
$$
D\mu\,=\,\prod_x d\Phi_1d\Phi_1^*d\Phi_2d\Phi_2^*dY_\mu dW_\mu^a.
$$
After transition to new variables the volume of gauge group $d\Omega$ is separated out
explicitly, $D\mu = d\Omega \,D\mu_{WS}$, leaving the gauge invariant part \cite{Faddeev:2008qc},
$$
D\mu_{WS}\,=\,\prod_x \rho^3d\rho\, dZ_\mu dW_\mu^+dW_\mu^-dA_\mu,
$$
where $\rho$ is the dilaton field, which unlike ordinary fields enters the measure
with local factor. The similar local factor occurs in the functional measure
for pure gravity, making it reparametrization invariant (for dimension four)\footnote{Here
the Euclidean signature is assumed for the metric tensor.},
$$
D{\cal G}\,=\,\prod_x\prod_{\mu\le \nu}{\cal G}^{\frac 52}d{\cal G}^{\mu\nu},
~~~{\cal G} = det \,{\cal G}_{\mu\nu}.
$$
To ensure the reparametrization invariance of the effective gravity in the presence
of matter fields the measure should be of the form
$$
D\mu_{eff}\,=\,D{\cal G}\prod_x {\cal G}^{-2}dZ_\mu dW_\mu^+dW_\mu^-dA_\mu,
$$
however the power of local factor differs from that in $D\mu_{WS}$.
In addition, the reparametrization invariance is violated by the Higgs mass term
in the action, $\mu^2|\Phi|^2=m^2\mu^2{\cal G}_{\nu\nu}/4$. This complicates the direct
match between electroweak and gravity theories beyond classical level.

We try to establish the connection starting from the completely covariant functional
integral,
$$
Z\,=\,\int D\mu_{eff} e^{-S_{WS}({\cal G}) -\frac 1\gamma W^2({\cal G})},
$$
with the action $S_{WS}$ corresponding to Lagrangian (\ref{LWS}) without Higgs mass term.
The square of Weyl tensor added to the action,
$W^{\,2}({\cal G})=\sqrt{{\cal G}}W_{\mu\nu\lambda\sigma}^2$, enforces the metric
to conformally flat form for $\gamma \to 0$ ($W_{\mu\nu\lambda\sigma}=0$ for conformally
flat metric)\cite{Chernodub:2007bz,Chernodub:2008rz}. One should stress the point that
the limit $\gamma \to 0$ actually singles out the metrics, which can be brought to the form
(\ref{rh}) after coordinate transformation. To get the form (\ref{rh}) precisely
the coordinates have to be fixed by four (for $D=4$) gauge conditions,
\begin{equation}
\label{eta}
\delta\bigl[\eta({\cal G}^{\mu \nu})-\ell\bigr]\,=\,\delta({\sqrt{{\cal G}}\cal G}^{11}-\ell)\,
\delta({\sqrt{{\cal G}}\cal G}^{22}-\ell)\,\delta({\sqrt{{\cal G}}\cal G}^{33}-\ell)\,
\delta({\sqrt{{\cal G}}\cal G}^{44}-\ell),
\end{equation}
with $\ell(x) = \rho^2(x)/m^2$.
After this the integral over ${\cal G}^{\mu\nu}$ turns actually into the integral over
coordinate transformations preserving the metric ${\cal G}_{\mu\nu}=\ell \delta_{\mu\nu}$,
so that
\begin{eqnarray}
\label{Zell}
Z(\ell)\,&=&\,\int D\mu_{eff} e^{-S_{WS}({\cal G})-\frac 1\gamma W^2({\cal G})}
\delta\bigl[\,\eta({\cal G}^{\mu \nu})-\ell\bigr] \\
&=&\,\frac 1{\Delta(\ell)}\,\int dZ_\mu dW_\mu^+dW_\mu^-dA_\mu e^{-S_{WS}(\ell)}.\nonumber
\end{eqnarray}
The function
$$
\frac 1{\Delta(\ell)}\,=\,\int D{\cal G}\prod_x {\cal G}^{-2}
e^{-\frac 1\gamma W^2({\cal G})}\,\delta\bigl[\eta({\cal G}^{\mu \nu})-\ell\bigr]
$$
involves the integral over residual for $\gamma \to 0$ coordinate transformations.
Introducing invariant functional measure $D\Omega$ on the group of coordinate
transformations, ${\cal G}^{\mu \nu} \to {\cal G}_{\,\Omega}^{\mu \nu}$, we can rewrite
it as
\begin{equation}
\label{FP}
\frac 1{\Delta(\ell)}\,=\,\frac 1{J(\ell)}\int D\Omega\,
\delta\bigl[\eta({\cal G}_{\,\Omega}^{\mu \nu})-\ell\bigr]\,=\,\frac 1{J(\ell)}
\frac 1{\Delta_{\rm FP}(\ell)},
\end{equation}
where $\Delta_{\rm FP}(\ell)$ is the standard Faddeev-Popov determinant for gravity,
function $J(\ell)$ comes from Jacobian relating the measures $D{\cal G}\prod_x {\cal G}^{-2}$
and $D\Omega$.

Equation (\ref{Zell}), (\ref{FP}) allows to connect the functional integrals for electroweak
theory and gravity,
\begin{eqnarray}
Z_{WS}\,&=&\,\int D\mu_{WS}e^{-S_{WS}(\ell)-m^2\mu^2\ell} \nonumber \\
&=&\,\int D\ell\,\ell^{\frac 32}\,\Delta(\ell)e^{-m^2\mu^2\ell}
\int D\mu_{eff} e^{-S_{WS}({\cal G}) -\frac 1\gamma W^2({\cal G})}
\delta\bigl[\eta({\cal G}^{\mu \nu})-\ell\bigr].\nonumber
\end{eqnarray}
Here we denote through $m^2\mu^2\ell=\int d^4x\mu^2|\Phi|^2$ the part of the action due to Higgs
mass term, $\ell^{\frac 32} \sim \rho^3$ stands for the local factor in the integration
measure $D\mu_{WS}$. Using the fact that $\Delta_{\rm FP}({\cal G}^{\mu \nu})=
\Delta_{\rm FP}(\ell)$ in the integral over metrics we finally have ($\gamma \to 0$)
\begin{eqnarray}
\label{Zgr}
Z_{WS}\,&=&\,\int D\ell\,J(\ell)\ell^{\frac 32}e^{-m^2\mu^2\ell} Z_{gr},  \\
Z_{gr}\,&=&\,\int D\mu_{eff} e^{-S_{WS}({\cal G}) -\frac 1\gamma W^2({\cal G})}
\Delta({\cal G}^{\mu \nu})\,\delta\bigl[\eta({\cal G}^{\mu \nu})-\ell\bigr]. \nonumber
\end{eqnarray}
The $Z_{gr}$ integral looks like a standard functional integral for gravity
interacting with matter fields written in the particular gauge (\ref{eta}).
According to general treatment $Z_{gr}$  does not depend on the gauge fixing parameter
$\ell$ (at least perturbatively), therefore the integral over $\ell$ with any weight
functional results into general normalization only. Thus we can conclude that
electroweak theory is equivalent at quantum (perturbative) level to a gravity
theory taken in a special gauge. Furthermore, the Higgs mass term in the action
along with the local factor in the measure $D\mu_{WS}$ and function $J(\ell)$
are interpreted as a particular weight functional
($\sim J(\ell)\ell^{\frac 32}\,e^{-m^2\mu^2\ell}$) in the chosen gauge.

Thus although in general case the Higgs mass term violates the complete covariance
of gravity theory we can treat this violation as a choice of special gauge. Only
after imposing this gauge and fixing very special form of the weight functional
the effective gravity becomes equivalent to the electroweak theory.

There are two possible ways to address the electroweak/gravity connection.
One way is to carry out calculations in the gravity under the gauge (\ref{eta})).
This way we may reproduce all the well known Standard Model results. On another hand it looks
attractive to take the "gravity" seriously and to consider the electroweak theory in the form
which is gauge (reparametrization) invariant.

A suitable method is to deal with those quantities in the electroweak theory which are gauge
(reparametrization) invariant with respect to the gravity. Any reasonable for the gravity gauge
can be imposed in this case with output valid for the electroweak theory. Besides, the output
value do not depend on the Higgs mass parameter $\mu^2$ in the Lagrangian (\ref{EW}).

The last method can be applied to $S$-matrix (in Minkowski space),
supposing its existence and reparametrization invariance in the gravity with
asymptotically flat spacetime boundary conditions. It allows to find scattering
amplitude directly in the effective gravity on the basis of conventional perturbation theory.

An interesting possibility caused by the universality of the gravitational interaction
is the direct coupling of the Higgs-dilaton to the photon.

\section{Photons pair production}
We take the process of photons pair production off $Z$-boson,
$e^+ + e^- \to Z \to Z + \gamma + \gamma$ shown in the Born order.

\begin{center}
\includegraphics[scale = 0.7]{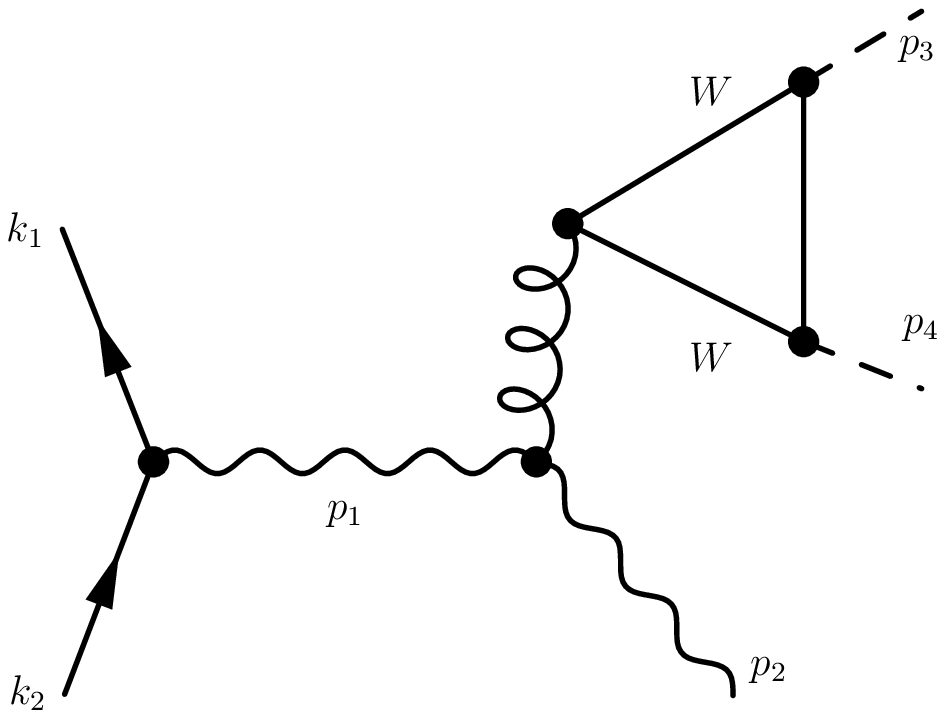}
\end{center}
{\sf
Photon pair production. $p_1$ and $p_2$ are $Z$-bosons momenta, curly line
denotes the dilaton, $p_{3,4}$ are the final photons momenta.}

\bigskip

To estimate the cross section of this reaction is interesting since:\\
i)this process demonstrates the universality of our electroweak gravity\\
ii) not excluded that the LEP2 data are already exclude (or confirm)
this 'gravity-like' approach.

From the effective gravity viewpoint this process occurs in the lowest order
through the virtual dilaton exchange. Although it looks rather simple there are two
difficulties to deal with. First, the coupling of the gravity to the matter fields
is universal and given by the energy-momentum tensor $\theta_{\mu\nu}$. In contrast
to graviton the dilaton feels the trace part of $\theta_{\mu\nu}$ while the photon
energy-momentum tensor is traceless. It makes scattering impossible at
the Born level, the interaction should be mediated by loops.

Second problem comes from the fact that virtual $Z$-boson is produced in $e^+e^-$
fusion, whereas the effective gravity describes pure bosonic sector only. To avoid
this difficulty we treat the process using unitarity, which is assumed to hold for
the whole amplitude. Having cut the diagram through virtual $Z$-boson line,
we left with two mass shell amplitudes: $e^+e^- \to Z$ and $Z \to Z + \gamma + \gamma$.
The former amplitude is really a standard electroweak neutral current vertex,
while the latter amplitude has the form suitable to apply
the effective gravity\footnote{In fact, we deal with kinematically allowed process
$Z+Z \to \gamma + \gamma$ and use crossing symmetry to get the amplitude required.}.
After this the full amplitude is recovered by the dispersion relation with respect
$Z$-boson invariant mass, which is simply amount to plugging the denominator of $Z$-boson
propagator into obtained expression. This trick is similar to those frequently used
in helicity based calculations \cite{MHV}.

We start first from the case, where $h \to \gamma +\gamma$ transition goes through
$W$-boson loop.
The dilaton vertices $ZZ \to h$, $h \to W^+W^-$, $W^+W^- \to \gamma$ as well as
the massive $W$-boson propagator can be immediately read off from the matter
Lagrangian (\ref{matter lagr}). The loop has been calculated by standard dispersion
relation taking the discontinuity with respect to the photon pair invariant energy,
$S_{\gamma\gamma}=(p_3+p_4)^2$ and subtraction point at $S_{\gamma\gamma}=0$. The latter choice is dictated by
the absence of real dilaton and photons interaction.

The dilaton propagator is extracted from the quadratic part of the gravity action,
$$
{\cal L}_{gr}\,=\,\sqrt{-{\cal G}}\bigl[-M_p^2R+\frac 1\gamma W_{\mu\nu\lambda\sigma}^2
\bigr],
$$
with the second term enforcing the metric to conformally flat form for $\gamma \to 0$.
Taking the metric as
$$
g^{\mu\nu}\,=\,\eta^{\mu\nu}\,+\,\frac 1{M_p^2}h^{\mu\nu}
$$
and adding to the action a gauge-fixing term,
\begin{eqnarray}
{\cal L}_{gf}\,&=&\,-\frac 14 \alpha\, h^{\mu\nu}
\bigl[\eta_{\lambda\sigma}\eta_{\mu\nu}\,\partial^2
+ \eta_{\nu\sigma}\,\partial_\lambda \partial_\mu
+ \eta_{\mu\sigma}\,\partial_\lambda \partial_\nu
- 2 \eta_{\mu\nu}\,\partial_\lambda \partial_\sigma
- 2 \eta_{\lambda\sigma}\, \partial_\mu \partial_\nu \nonumber \\
&&+ \eta_{\lambda\nu}\,\partial_\mu \partial_\sigma
+ \eta_{\lambda\mu}\,\partial_\nu \partial_\sigma \bigr]\,h^{\lambda\sigma} \nonumber
\end{eqnarray}
arranged to produce harmonic (de Donder) gauge,
$$
\partial_\mu h^{\mu \nu} - \frac 12\partial_\nu h^{\mu \mu} = 0,
$$
we arrive at the propagator ($\gamma=0$)
\begin{eqnarray}
G_{\mu \nu \lambda \sigma}(k)\!&=&\!\frac 1{6 k^6 \,\alpha}
\bigl[k^4 \eta_{\lambda\sigma} \eta_{\mu\nu} \alpha
+ 6 k^2 k_\lambda k_\mu \eta_{\nu\sigma}
+ 6 k^2 k_\lambda k_\nu \eta_{\mu\sigma}
+ 2 k^2 k_\lambda k_\sigma \eta_{\mu\nu} \alpha \nonumber \\
&&+ 2 k^2 k_\mu k_\nu \eta_{\lambda\sigma} \alpha
+ 6 k^2 k_\mu k_\sigma \eta_{\lambda\nu}
+ 6 k^2 k_\nu k_\sigma \eta_{\lambda\mu}
+ 4 k_\lambda k_\mu k_\nu k_\sigma \alpha\bigr].\nonumber
\end{eqnarray}

One has to emphasize at this point that we assume the $S$-matrix element for
$Z \to Z + \gamma + \gamma$ subprocess to be reparametrization invariant within
effective gravity. It is this property that allows us to choose the convenient
gauge and omit the Higgs mass term as has been discussed in the previous section.

Combining the dilaton exchange amplitude with the amplitude produced by the relevant
piece of neutral current, where we neglect the lepton masses,
$$
J_Z^\mu\,=\,\frac {g}{\cos\theta_W}\bigl[\,\overline{e}\gamma^\mu
\bigl(\sin^2\theta_W-\frac 14\bigr)e\,
+\,\frac 14\,\overline{e}\gamma^\mu\gamma^5 e\bigr],
$$
($\theta_W$ is Weinberg angle),
we get the imaginary part through which the final amplitude is restored.
The cross section is summed up over two photons' polarizations, over three polarizations
of outgoing $Z$-boson and averaged over incoming massless electron helicities. Finally it
is integrated over phase volume with the invariant energy of the photon pair, $S_{\gamma\gamma}=(p_3+p_4)^2$,
fixed. The resulting one-loop cross section in the leading order both in $S_{\gamma\gamma}$
and the lepton pair invariant energy, $S=(p_1+p_{\,2})^2$, reads
$$
\frac{d\sigma}{d S_{\gamma\gamma}}\,=\,\frac{g^2}{512\pi^3}
\frac{1-4\sin^2\theta_W+8\sin^4\theta_W}{\cos^2\theta_W}
\frac{\alpha_{em}^2}{16\pi^2}\,\frac{M_Z^2}{1152 M_p^4}\,\frac 1S.
$$

There is a way to take into account also the contribution of fermion loop mediating
$h \to \gamma+\gamma$ transition. Indeed, the fermion loop contribution is rather
universal in QED and given by the well-known anomaly of energy-momentum tensor,
\begin{equation}
\label{anomaly}
\theta_\mu^\mu\,=\,\frac{\beta(e)}{2e} F_{\mu\nu}F^{\mu\nu},
\end{equation}
relating the trace value to the electrodynamic $\beta$-function. On the other hand,
this part of energy-momentum tensor provides the vertex for dilaton-photon coupling.
Putting it together with pure boson loop found above we arrive at
$$
\frac{d\sigma}{d S_{\gamma\gamma}}\,=\,\frac{g^2}{512\pi^3}
\frac{1-4\sin^2\theta_W+8\sin^4\theta_W}{\cos^2\theta_W}
\biggl[\frac{\alpha_{em}}{2\pi}-\frac{\beta(e)}{e}\biggr]^2\frac{M_Z^2}{1152
M_p^4}\, \frac 1S,
$$
where $\beta(e)=n_f\frac{e^3}{12\pi^2}$ is one loop beta-function of QED with
$n_f$ different fermions.

Recalling that all parameters are expressed through a single
scale in the effective gravity (\ref{mscale}),
$$
M_p^2\,=\,\frac 13 M_Z^{\,2}\,\frac{\cos^2\theta_W}{g^2},
$$
we finally have for the one loop $e^+e^- \to Z +\gamma+\gamma$
cross section
\begin{equation}
\label{fin}
\frac{d\sigma}{d S_{\gamma\gamma}}\,=\,\frac 98 \frac{\alpha_{em}^5}{16\pi^2}\,
\frac{1-4\sin^2\theta_W+8\sin^4\theta_W}{\sin^6\theta_W\cos^6\theta_W}\,
\bigl(1-\frac 23\,n_f\bigr)^2\,
\frac{1}{1152 M_Z^2}\,\frac 1S.
\end{equation}
In case of fermions with various charges, $e_f=e_Qe$, $1-\frac 23\,n_f \to 1-\frac 23\sum e_Q^2$.

Numerically the obtained cross section is extremely small,
$\sigma \simeq 5\cdot 10^{-9}$fb for $n_f=0$.
Therefore we had no chance to observe the corresponding process at LEP2 where
the integrated luminosity per experiment was of the order of 1~fb$^{-1}$.

It could be of interest to compare above results with the cross section of the same process
but with conventional massive scalar Higgs and standard electroweak vertices.
The similar calculation with Higgs coupled to photons via $W^\pm$ triangle
gives for $S,S_{\gamma\gamma} \gg m_H^2$
\begin{equation}
\label{standard}
\frac{d\sigma}{d S_{\gamma\gamma}}\,\simeq\,\frac{3}{64} \frac{\alpha_{em}^5}{16\pi^2}\,
\frac{1-4\sin^2\theta_W+8\sin^4\theta_W}{\sin^6\theta_W\cos^2\theta_W}\,
\frac{M_Z^2}{S S_{\gamma\gamma}^2}\,\log^4\frac{S_{\gamma\gamma}}{M_W^2}.
\end{equation}
One more possible contribution occurs when the process goes through fermion loop.
Taking the light quarks with masses $m_Q$ we get in the lowest in $m_Q$ order an additional
term to the cross section (\ref{standard}),
\begin{equation}
\label{fermion}
\frac{d\sigma_{f}}{d S_{\gamma\gamma}}\,\simeq\,\sum_Q\frac{1}{32}
\frac{\alpha_{em}^5}{16\pi^2}\,
\frac{1-4\sin^2\theta_W+8\sin^4\theta_W}{\sin^6\theta_W\cos^2\theta_W}\,
\frac{m_Q^2e_Q^2}{S S_{\gamma\gamma}^2}\,\log^2\frac{S_{\gamma\gamma}}{M_W^2}
\log^2\frac{S_{\gamma\gamma}}{m_Q^2}.
\end{equation}
Actually, it results from interference between $W^\pm$ and quark triangles
in the amplitude squared, the sum is taken over different fermions contributions
whose masses and charges are $m_Q$ and $e_Q$, the quark loop being proportional
to $m_Q$ with an extra $m_Q$ appearing due to standard fermion-Higgs vertex.

The cross section has the same order in this case, though it is lesser
than what follows from Eq.(\ref{fin}) for $n_f=0$, the $W^\pm$ triangle,
Eq.(\ref{standard}), yields $\sigma \simeq 1.7\cdot 10^{-9}$fb for $\sqrt{S}=14$~TeV.
The reason for the discrepancy comes from the different asymptotic of the results
for dilaton and conventional Higgs. According to the expressions (\ref{standard}),
(\ref{fermion}) the total cross section $e^+ + e^- \to Z + \gamma + \gamma$ decreases
with the invariant energy, $\sigma \sim 1/S$. On the contrary, the expression (\ref{fin})
leads to the constant cross section, which, therefore, exceed the value predicted
by Eqs.(\ref{standard}), (\ref{fermion}) for large enough $S$. There is an additional
difference between the effective gravity and standard model expressions.
The term including fermion loop (\ref{fermion}) vanishes for zero fermion mass
while the cross section evaluated through the dilaton exchange
is finite for massless fermions, depending on their number $n_f$ only.

The cross sections can be compared for the small energies too. Near threshold of
the reaction, $\delta S = S - (M_Z+M_{\gamma\gamma})^2 \to 0$, assuming $S_{\gamma\gamma}=
M_{\gamma\gamma}^2 \ll M_Z^2$, the dilaton exchange process yields in the lowest
order in $M_{\gamma\gamma}/M_Z$
\begin{eqnarray}
\frac{d\sigma}{d S_{\gamma\gamma}}\,&\simeq&\,\frac 1{2048}\frac{\alpha_{em}^5}{16\pi^2}\,
\frac{1-4\sin^2\theta_W+8\sin^4\theta_W}{\sin^6\theta_W\cos^6\theta_W}\,
\frac 1{S}\frac {S^2-2M_Z^4}{(S-M_Z^2)^2} \nonumber \\
\label{dilat0}
&\times&\,\frac 1{M_Z^3}\sqrt{\frac{M_{\gamma\gamma}}{M_Z}\delta S}\,
\bigl(1+\frac 83 \sum e_Q^2\bigr)^2,
\end{eqnarray}
(the opposite sign with respect to the high energy limit (\ref{fin}) in front of $e_Q^2$
in the last bracket is due to the $W^\pm$ triangle amplitude behavior for small
$S_{\gamma\gamma}$). It gives for the total cross section $\sigma \sim 10^{-9}$~fb
for $\sqrt{S}=200$~GeV and for $n_f=0$.

The cross section value for the standard massive Higgs attached to the $W^\pm$
triangle is estimated for $S_{\gamma\gamma}\ll M_Z^2 \ll m_H^2$ as
\begin{eqnarray}
\frac{d\sigma}{d S_{\gamma\gamma}}\,&\simeq\,&\frac {25}{64}\,\frac{\alpha_{em}^5}{16\pi^2}\,
\frac{1-4\sin^2\theta_W+8\sin^4\theta_W}{\sin^6\theta_W\cos^2\theta_W}\,
\frac 1{S}\frac 1{(S-M_Z^2)^2}\frac{M_Z^5 M_{\gamma\gamma}^4}{m_H^4M_W^4}
\nonumber \\
\label{higgs0}
&\times&\sqrt{\frac{M_{\gamma\gamma}}{M_Z}\delta S},
\end{eqnarray}
Like the previous expression (\ref{dilat0}), this formula is derived in the lowest order
in $M_{\gamma\gamma}/M_Z$. Although equations(\ref{dilat0}) and (\ref{higgs0}) look different,
their numeric values are rather close. Taking $\sqrt{S}=200$~GeV and $m_H=200$~GeV we get
for Eq. (\ref{higgs0}) $\sigma \simeq 10^{-9}$~fb.

Thus, summarizing the above results and limiting cases, the cross sections due to dilaton and
standard massive Higgs exchange are almost the same near threshold of
the $e^+ + e^- \to Z + \gamma + \gamma$ reaction. The standard Higgs mediated cross section
grows more rapidly with the energy and exceeds the dilaton one, which
tends to a constant at the energies greater than the typical mass scale
of the effective gravity, $S\gg M_p^2\sim 5$~TeV$^2$. However starting from the energies
$\sqrt{S} \simeq 100$~GeV the Higgs exchange cross section decreases, becomes equal
to the dilaton cross section at $\sqrt{S} \simeq 8$~TeV and goes to zero for asymptotically
large energies.

\section{Conclusion}

There are two main ingredients in the electroweak theory treatment in terms of the effective
gravity. The masses of vector bosons are introduced without appealing to symmetry breaking
mechanism through Higgs field condensate. The Higgs particle is treated as massless dilaton.
The term with Higgs mass in the original electroweak Lagrangian violates the reparametrization
invariance of the effective gravity. However this term can be interpreted as a part
of gauge-fixing functional for a special gauge chosen in the gravity. If the quantities
we are interested in the electroweak theory are gauge invariant from the viewpoint
of effective gravity (do not depend on the coordinate system), the perturbative relations
(\ref{Zgr}) allow to obtain them directly in the effective gravity taking any convenient gauge.

Note that though the Higgs mass term $\mu^2$ in the Lagrangian (\ref{EW}) does not affect
the final result it could in principle depend on the Higgs self-interaction term $\lambda$,
playing in gravity the role of cosmological constant. However $S$-matrix becomes subtle
since the gravity is no more asymptotically flat in this case.

Recall that from the experimental viewpoint at the moment we know
nothing about the self-interaction of Higgs boson. The conventional $\lambda|\Phi|^4$ potential
is just the simplest possibility. Thus we may discuss another forms of the Higgs boson
potential.

The one loop cross section $e^+ + e^- \to Z + \gamma + \gamma$ obtained in the effective
gravity though very small shows the quantitative and qualitative difference with
the cross section found in the standard electroweak theory. The discrepancy seems to come
about due to asymptotically flatness of the effective gravity we have dealt with,
that is $\lambda=0$, which is not the case for the standard theory. Another reason probably
lies in the fact that employing effective gravity approach we thereby reformulate in somewhat
manner the perturbation theory, so that there is no order by order correspondence with those
used in the standard model.

Strictly speaking we have to worry about the radiative corrections.
The global fit to electroweak data in the standard Weinberg-Salam form indicates that
the mass of the Higgs boson should be of the order of 100~GeV. In the case of a massless
Higgs-dilaton the infrared behavior should be sensitive to space-time structure at infinity.
The corresponding effect arises from the "curvature of the space" resulting
from scalar field self-interaction. That is why it could be of interest to get further insight
into effective gravity approach, in particular, to study high orders of perturbative expansion.

\end{document}